\documentclass[journal]{IEEEtran}%
\IEEEoverridecommandlockouts

\usepackage{soul}
\usepackage{cite}

\usepackage{diagbox}
\usepackage{amsmath,amssymb,amsfonts}
\usepackage{amsthm,bm}
\usepackage{mathrsfs}
\usepackage{colortbl}
\usepackage{hhline}
\usepackage{tabularx}
\usepackage{enumerate}
\usepackage{booktabs}
\usepackage{booktabs}

\usepackage{multicol}
\usepackage{graphicx}
\usepackage{subfigure} 
\usepackage{textcomp}
\usepackage{url}
\usepackage{stfloats}
\usepackage{mathtools}
\usepackage{cuted}
\usepackage{array}  
\usepackage{makecell}
\usepackage {threeparttable}
\usepackage{tablefootnote}
\usepackage{multirow}
\usepackage{colortbl}
\usepackage{setspace}
\def\BibTeX{{\rm B\kern-.05em{\sc i\kern-.025em b}\kern-.08em
    T\kern-.1667em\lower.7ex\hbox{E}\kern-.125emX}}

\usepackage{stackengine}
\usepackage{enumerate}
\usepackage{booktabs}
\usepackage{subfig}
\usepackage{multicol}
\usepackage[noend]{algpseudocode}
\usepackage[table,xcdraw]{xcolor}
\usepackage{pgfplots}
\usepackage{pgfplotstable}
\usepackage[ruled,linesnumbered]{algorithm2e} 
\SetKwRepeat{Do}{do}{while}
\definecolor{green}{RGB}{0, 176, 80}
\usepackage{setspace}
\usepackage{color}
\usepackage{tabularray}
\definecolor{Iron}{rgb}{0.811,0.815,0.815}
\makeatletter
\def\BState{\State\hskip-\ALG@thistlm}
\makeatother

\algnewcommand\algorithmicforeach{\textbf{for each}}
\algdef{S}[FOR]{ForEach}[1]{\algorithmicforeach\ #1\ \algorithmicdo}
\begin{document}

\title{Rethinking Generative Semantic Communication  for \\ Multi-User Systems with Large Language Models}
\author{Wanting Yang, Zehui Xiong, \textit{Senior Member, IEEE}, Shiwen Mao, \textit{Fellow, IEEE}, Tony Q. S. Quek, \textit{Fellow, IEEE}, \\ Ping Zhang, \textit{Fellow, IEEE}, Mérouane Debbah, \textit{Fellow, IEEE},  Rahim Tafazolli, \textit{Fellow, IEEE}  \thanks{ Wanting Yang, Zehui Xiong, Tony Q. S. Quek are with the Pillar
of Information Systems Technology and Design, Singapore University of
Technology and Design, Singapore (e-mail: wanting\_yang@sutd.edu.sg; zehui\_xiong@sutd.edu.sg; tonyquek@sutd.edu.sg) Shiwen Mao is with the Department of Electrical and Computer Engineering, Auburn University, Auburn
36830, USA (e-mail: smao@ieee.org); 
P. Zhang is with the School of Information and Communication Engineering, Beijing University of Posts and Telecommunications, and also with
the State Key Laboratory of Networking and Switching Technology, Beijing
100876, China (email: pzhang@bupt.edu.cn).
M. Debbah is with  KU 6G Research Center, Department of Computer and Information Engineering, Khalifa University, Abu Dhabi 127788, UAE (email: merouane.debbah@ku.ac.ae) and also with CentraleSupelec, University Paris-Saclay, 91192 Gif-sur-Yvette, France. 
Rahim Tafazolli is with the Institute
for Communication Systems (ICS), 5GIC \& 6GIC, University of Surrey,
Guildford, Surrey, GU2 7XH, U.K. (email:
r.tafazolli@surrey.ac.uk).

}
}

\makeatletter
\setlength{\@fptop}{0pt}
\makeatother

\maketitle

\vspace{-1.8cm}
\begin{abstract}
The surge in connected devices in 6G with typical complex tasks requiring multi-user cooperation, such as smart agriculture  and smart cities, poses significant challenges to unsustainable traditional communication. Fortunately, the booming artificial intelligence technology and the growing computational power of devices offer a promising 6G enabler: semantic communication (SemCom). However, existing deep learning-based SemCom paradigms struggle to extend to multi-user scenarios due to its increasing model size with the growing number of users and its limited compatibility with complex communication environments. Consequently, to truly empower 6G networks with this critical technology, this article rethinks generative SemCom for multi-user system  and proposes a novel framework called ``M-GSC" with the large language model (LLM) as the shared knowledge base (SKB). The LLM-based SKB plays three critical roles, that is, complex task decomposition, semantic representation specification, and semantic translation and mapping, for complex tasks, spawning a series of benefits such as semantic encoding standardization and semantic decoding personalization. Meanwhile, to enhance the performance of M-GSC framework, we highlight three optimization
strategies unique to this framework: extending the LLM-based SKB into a multi-agent LLM system, offloading semantic encoding and decoding, and managing communication and computational resources. Finally, a case study is conducted to demonstrate the preliminary validation on the effectiveness of the M-GSC framework in terms of efficient decoding offloading.

\end{abstract}


\newtheorem{definition}{Definition}
\newtheorem{lemma}{Proposition}
\newtheorem{theorem}{Theorem}

\newtheorem{property}{Property}

\vspace{0cm}
\section{Introduction}
\label{sec: Intro}
The future network is envisioned to enable data-intensive interactions among humans, machines, and environments. The escalating data demand is straining the increasingly limited available radio resources. This creates a significant challenge for complex communication tasks that rely on multi-user cooperation.
Much like the other fields benefiting from advancements of artificial intelligence (AI), semantic communication (SemCom)  invigorated by AI is on the brink of tight integration into 6G. 

Akin to enhancing inter-human communication efficiency, SemCom achieves the reduction of the data amount to be transmitted via task-oriented semantic parsing and extraction  before transmission.
However, it simultaneously increases computational demands on end devices, making it challenging to generalize SemCom across devices with varying capabilities.
Thankfully, given the widespread acceptance of edge computational offloading, integrating the edge into the  SemCom is a necessary and  natural  progression. Nonetheless, despite its potential, SemCom as a service--envisioned as a major potential  functionality provided by seamless  end-edge cooperation--has not garnered much interest. 

More critically, existing SemCom frameworks are primarily optimized for point-to-point scenarios and fixed small-scale communication groups. They are not yet equipped to handle dynamic  multi-user environments in 6G. Specifically, the most dominant deep-learning (DL)-based SemCom frameworks are characterized by two primary features:  \textit{end-to-end training}   and \textit{analog transmission}\footnote{In the SemCom paradigm of analog transmission, the encoder's output symbols are directly fed into the channel via amplitude modulation, without a quantization process.}. During the end-to-end training, to ensure the informative quality of transmitted semantic information, the values of the elements output by the semantic encoder tend to be unconstrained. As a result, the values exhibit potentially large variations and an unknown range.  If digital transmission is applied, maintaining precision would require lengthy bit sequences for each element~\cite{guo2023digital}, to prevent semantic ambiguity during quantization. Given the high number of output elements, this weakens the robustness of SemCom in low signal-to-noise ratio (SNR) conditions. 
Meanwhile, although DL-based SemCom demonstrates promising performance and remarkable efficiency in \textit{short-range} communication, it lacks error correction. This makes it highly vulnerable to noise accumulation, restricting its applicability in long-distance multi-hop scenarios.


Simultaneously, the end-to-end training manner inherently treats the semantic encoders and decoders of all users as a unified whole. When integrated with advanced communication techniques, such as trainable quantization module~\cite{guo2023digital} and dynamic channel state information fusion~\cite{10559783}, the training overhead is further enhanced.
To address the computational limitations of large-scale model training, preliminary works have integrated edge computing into SemCom frameworks to offload training tasks \cite{xie2020lite, yang2022semanticmaga}. However, such integration often idealistically assumes that semantic encoding and decoding can be entirely performed on end devices, despite significant feasibility challenges.  To this end, recent research on DL-based SemCom~\cite{guo2023digital}   proposes a trainable policy for semantic encoding offloading. 
Nevertheless, in scenarios with an expanding range of tasks, the task-specific training paradigm could necessitate an impractical number of unique semantic encoder and decoder pairs for each device. In this light, such a paradigm appears impractical for scaling to the complex and dynamic  communication scenarios such as extended reality and intelligent transportation anticipated in 6G. 

\begin{figure*}
    \centering
    \includegraphics[width=1\linewidth]{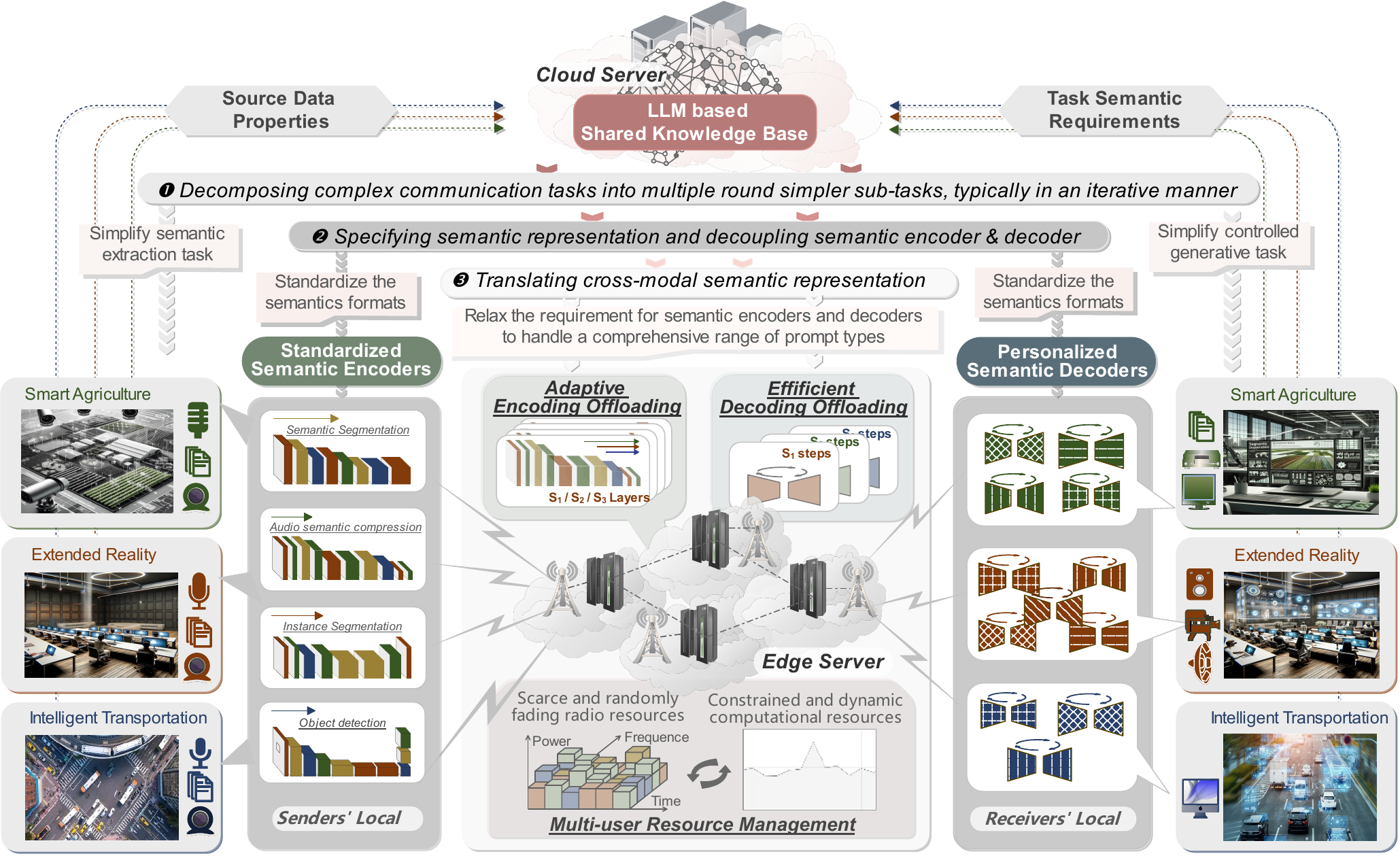}
\caption{Illustration of M-GSC framework within a cloud-edge-end network architecture.}
    \label{fig:overview}
\end{figure*}

Opportunely, as stated in our preliminary work~\cite{10577142}, the burgeoning field of Generative Artificial Intelligence (GAI), particularly diffusion models, shows great potential to enrich and extend the existing SemCom paradigm towards generative SemCom. It enables a shift from the black box of end-to-end training to the glass box approach, where the semantic encoder and decoder are decoupled by a given semantic representation and transformed into the prompt engineering and controlled generation, respectively. This facilitates the exploration of remarkable advancements in natural language processing and computer vision to enable  to enable cross-modal capabilities, which, in turn, achieve higher compression rates~\cite{raha2023generative}.
  However, the following advantages of generative SemCom, particularly for multi-user systems, remain largely overlooked. 
\begin{itemize}
    \item The glass-box nature of generative SemCom enhances the explainability of semantic information representations, such as natural language, semantic segmentation maps, and bounding boxes. This facilitates the creation of a consensus-based semantic knowledge space among multiple users. It allows users to systematically establish and adopt shared semantic representations, ensuring consistency in semantic understanding and enhancing communication efficiency.
    \item The decoupling paradigm enables standardized semantic encoding and personalized semantic decoding. By adopting a consensus set of semantic representations, general-purpose models like the Segment Anything Model (SAM) and YOLO can support multiple tasks. This approach minimizes the deployment redundancy. Meanwhile, diffusion-based semantic decoders can integrate user preferences through additional local prompts or fine-tuning techniques like DreamBooth. This facilitates customized applications effectively.
    \item The independent training of semantic encoders and decoders removes the constraint of channel conditions. Additionally, the controllability of semantic representation facilitates efficient quantization, exemplified by JPEG image encoding for semantic segmentation maps. These advantages make generative SemCom more compatible with existing digital transmission networks. As a result, it retains the noise and interference resistance of digital transmission while enabling the trained model to adapt to diverse channel environments.
\end{itemize}

Given its promising potential, we rethink a unified generative SemCom framework, with three new key considerations for  multi-user scenarios with diverse and complex tasks.
\begin{itemize}
    \item [C1]Generative SemCom requires a more sophisticated shared knowledge base (SKB). In DL-based SemCom, the \textit{raw data} themselves serve as the SKB, with neural networks relying on them to analyze and extract semantic information during training. In contrast, generative SemCom depends on the SKB to provide \textit{specialized knowledge} for targeted guidance in designing encoders and decoders.
    \item [C2]The inference process of large-scale semantic encoders and diffusion-based semantic decoders may result in higher computational demands and longer inference latency.
Addressing the computational power disparities across end devices is crucial for the widespread adoption of SemCom in multi-user scenarios and warrants further consideration. 
\item [C3] Extending SemCom to multi-user scenarios emphasizes the need for novel resource management  method. Unlike traditional content-agnostic networks, semantic-aware optimization can identify potential redundancy in transmitted data at the semantic level. This redundancy exists across multiple users, allowing for further reductions in communication overhead.
\end{itemize}


With the above in mind, we propose a novel  multi-user generative SemCom framework called ``M-GSC", which is tightly integrated with the cloud-edge-end architecture of the 6G network as shown in Fig.~\ref{fig:overview}. The contribution of our work has been highlighted as below.
\begin{itemize}
     \item For C1, we introduce a  large language model (LLM) based SKB for generative SemCom.  The chain-of-thought (CoT) capability of LLMs enables hierarchical semantic parsing for the targeted task, promoting adaptive sequential semantic extraction to   accommodate dynamic communication environments. Moreover, we propose a multi-LLM agent framework for enhanced SKB with greater robustness and adaptability. 
    \item In response to C2, we propose two novel offloading paradigms designed specifically for adaptive semantic encoding offloading and efficient semantic decoding. Moreover, addressing  C3, we further highlight tailored optimization strategies for the unique management of communication and computational resources in M-GSC.

    \item We conduct a case study using a simplified prototype of the M-GSC framework, with multiple YOLO based semantic encoders and   custom-trained diffusion-based decoders.  We demonstrate  the effectiveness of federated learning based efficient decoding offloading through experimental results. Additionally, we compare the quality of decoded data with DL-based SemCom benchmark to highlight its superiority.
\end{itemize}

The remainder of this paper is organized as follows. Section~\ref{sec:overvie} introduces  the design and architecture of M-GSC framework with the assistance of LLM-based SKB. Section~\ref{sec: edge-enhancement} presents the proposed optimization strategies for the M-GSC. In Section~\ref{sec: casestudy}, we conduct a case study, followed by the conclusion in Section~\ref{sec: con}.

\section{Framework Overview and Design Insights}
\label{sec:overvie}

In this section, we first introduce the motivation for using LLMs in SKB, then highlight their suitability by presenting key design insights of M-GSC. 

\subsection{Rational for Integrating LLM into M-GSC as SKB}
The SKB acts as a unique auxiliary module to enhances efficiency in SemCom. It is required to provide global background knowledge for semantic extraction and inference~\cite{ren2023knowledge}. As stated in C1, unlike the empirical datasets used in DL-based SemCom with end-to-end training manner, generative SemCom requires an SKB which goes beyond raw data to incorporate higher-level knowledge and even wisdom, enabling interpretable semantic extraction and inference. Recently, with the rapid proliferation of  LLMs, such as ChatGPT-4 and Deepseek, their remarkable and general semantic comprehension abilities have drawn widespread attention across various fields. This suggests that they have the potential to serve as the universal SKB in M-GSC. 

In fact, integrating knowledge from pre-trained LLMs into SemCom has garnered some academic attention, though it remains an emerging area of interest. In preliminary studies ~\cite{wang2024uses,zhao2024lamosc,jiang2023large}, LLMs are primarily used to provide side information to support training and enhance the performance of semantic encoders and decoders. Their black-box nature has limited their role to auxiliary tasks. The  most impressive abilities, such as  CoT reasoning and long-context understanding, remain largely underutilized. Hopefully, the decoupling of  semantic encoders and decoders   opens up the possibility for LLMs to intervene deeply in the SemCom system design.  The deployment of SKB is detailed in Section~\ref{sec: overview}, while Section~\ref{sec: SKB-agent} introduces a multi-LLM-agent system designed to enhance SKB with greater robustness and adaptability.

\begin{figure*}
    \centering
    \includegraphics[width=1\linewidth]{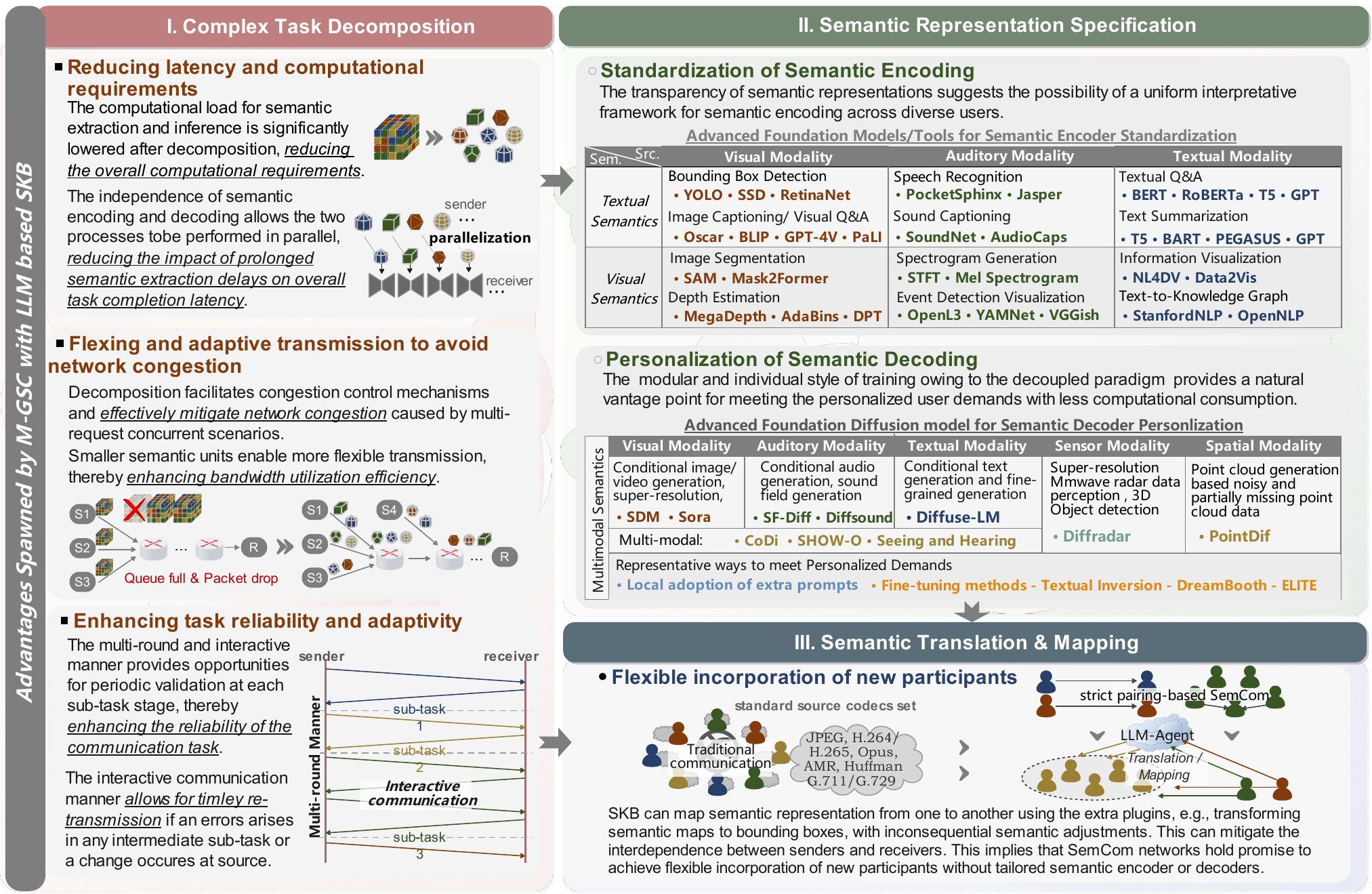}
    \caption{Summary of advantages spawned by integrating LLM-based agent into generative SemCom.}
    \label{fig:advantages}
\end{figure*}

\subsection{Architectural Overview and Insights for M-GSC}
\label{sec: overview}
The overview of M-GSC is shown in Fig.~\ref{fig:overview}. We assume that the LLM-based SKB can be deployed on distributed cloud servers, making it accessible to users in different geographic regions and the generalist model can serve various tasks. When the senders or receivers intend to
initiate a communication task, a request is first sent to the
LLM-based SKB. Then, the subsequent communication flow is sequentially outlined based on the three core functions of the SKB.
\subsubsection{Complex Task Decomposition}After the SKB receives a task request, LLMs  leverage the CoT reasoning capability to decompose the complex communication task into multiple sub-tasks with logical relationships and dependencies.  This
decomposition is based on the communication environment at
a large time scale.  As an example, the communication task for virtual scene construction task can be roughly split by ChatGPT-4 into:  \textit{basic environment $\to$ static buildings $\to$ dynamic elements $\to$ interactive objects $\to$ lighting and effects $\to$ texture mapping \& material assignment}. Compared to the complex task of directly generating a complete multi-modal virtual scene, a multi-round, interactive approach is much more manageable. More importantly, unlike manually predefined and fixed task decomposition methods, the SKB can dynamically adjust sub-tasks based on the actual communication network environment. This adaptability allows the system to offload more details to the generation process at the receiver when communication resources are limited, ensuring efficient task execution. 
\subsubsection{Semantic Representation Specification}Afterward, the LLM-based SKB identifies mutually agreed-upon semantic representations for each sub-task. 
For instance, in constructing basic environments and static buildings, suitable semantic representations may include semantic segmentation, depth maps, and partially masked point clouds, each offering different levels of detail and varying data sizes.  The SKB can select the most suitable semantic representation or the optimal combination of available representations, to maximize semantic accuracy while ensuring transmission reliability. In this context, the  senders can utilize multiple  standardized semantic encoders, e.g., SAM and YOLOs.  More foundation models are summarized in Fig.~\ref{fig:advantages}, which can be generalized across multiple task.  Meanwhile, 
 the receivers can deploy multiple diffusion models as semantic decoders based on the required modal data, each designed to generate different targets. As shown in Fig.~\ref{fig:advantages}, these models can support prompts corresponding to various forms of semantic representation. Furthermore, user-specific personalized decoders can be achieved by incorporating local personalized prompts or applying fine-tuning methods like DreamBooth, enhancing adaptability to personalized requirements.

\subsubsection{Semantic Translation \& Mapping}
If the sender or receiver lacks the semantic encoders or decoders for the designated  semantic representation, the LLM-based SKB can facilitate the semantic transformation and mapping. This enables SemCom networks to seamlessly incorporate a temporary participant without requiring extensive pre-training for tailored semantic encoders and decoders.
For example, transforming
semantic maps to bounding boxes requires the following series
of operations: segmentation masks $\to$ connected component
analysis $\to$ bounding box calculation, which can be achieved
by the SKB through CoT and extra plugins. 
Although effective, marginally, such
transformations may lead to the loss of semantic information
or the unwarranted  semantic content. 

Due to space limitations, more advantages spawned by integrating LLM-based SKB are detailed in Fig.~\ref{fig:advantages}.
\label{sec:IIB}
\begin{figure*}
    \centering
    \includegraphics[width=0.95\linewidth]{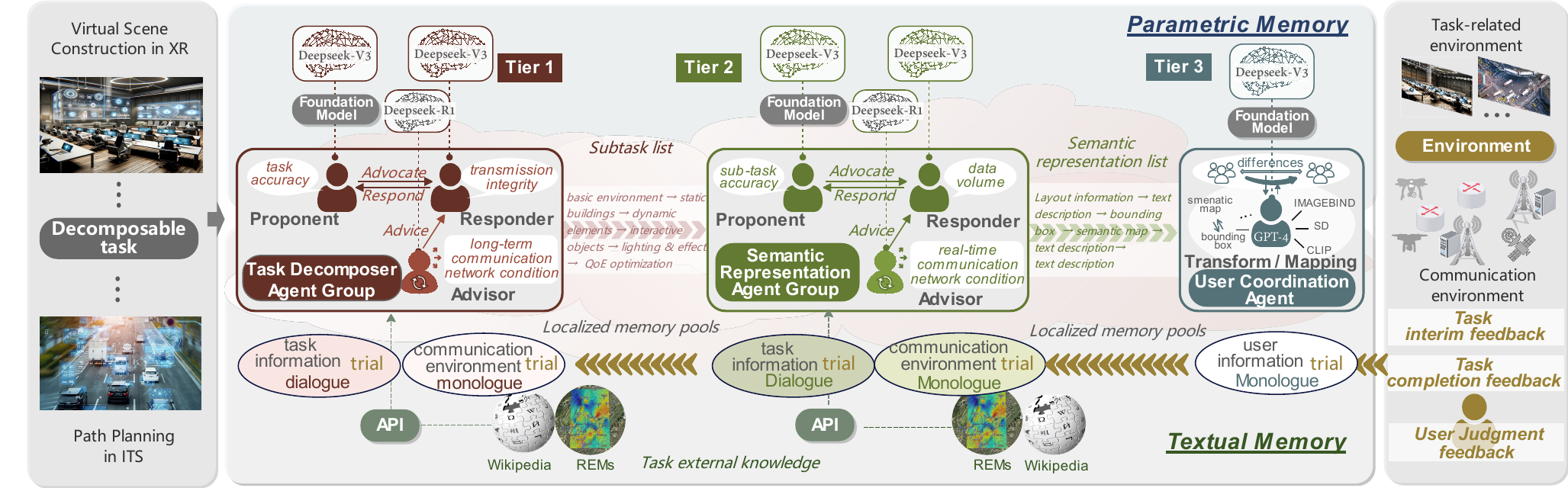}
    \caption{Illustration of LLM based-SKB multi-agent system design.}
    \label{fig:SKB}
\end{figure*}

\section{Optimization Strategies for M-GSC}
\label{sec: edge-enhancement}
In this section, we focus on the optimization strategies for the M-GSC framework, sequentially corresponding to the considerations C1–C3 stated in Section~\ref{sec: Intro}.

\subsection{Multi-LLM-agent System  for enhanced SKB}
\label{sec: SKB-agent}
Although LLMs' powerful reasoning capabilities hold great potential for optimizing and emulating human workflows, a single LLM cannot entirely eliminate the risk of hallucinations. To address this, we design a multi-agent system, inspired by the way humans collaborate to improve decision accuracy. Additionally, we introduce two trainable modules to further refine the system's expertise in communication-related tasks.

As shown in Fig.~\ref{fig:SKB}, the SKB follows a three-tier hierarchical cooperation structure. \textit{Tier 1, Task Decomposer}, breaks down complex tasks into sub-tasks and generates a corresponding semantic representation list for \textit{Tier 2, Semantic Representation}. This list then informs the user coordination decisions made by \textit{Tier 3, User Coordination}. The first two tiers each consist of three agents: \textit{Proponent, Responder, and Advisor}. The \textit{Proponent} and \textit{Responder} engage in a debate paradigm, defending their respective perspectives. In the \textit{Task Decomposer} tier, the Proponent prioritizes task performance, while the Responder focuses on transmission reliability, considering session latency requirements. The \textit{Advisor} supports the Responder by providing long-term communication insights to evaluate the sub-task list.
Similarly, in the \textit{Semantic Representation} tier, the Proponent ensures the accuracy of each sub-task, while the Responder aims to balance the data volume of semantic representations with the available transmission capacity. Here, the Advisor provides real-time communication conditions to assist the Responder.
Finally, the \textit{User Coordination} tier functions as a translator, converting semantic representations from the sender into a format compatible with the receiver if necessary. 

Upon receiving a request, task requirements and communication conditions—including antenna configurations, base station deployment, and core network topology—are provided to Tier~1 and Tier~2. Meanwhile, user-specific details for the sender and receiver, such as available semantic encoders and decoders, are sent to Tier~3.
To ensure each agent accesses only relevant information and remains focused, five localized memory pools store historical dialogues or monologues, task details, and user information. Since overall performance depends on all modules, historical environmental feedback and user judgment feedback are also recorded in each pool. Each agent then leverages its own memory pool as a valuable prompt to guide future decision-making.
Moreover, the Proponents, Responders, and User Coordination Agent can be implemented using well-trained, general-purpose models with strong reasoning capabilities for knowledge-intensive tasks, such as DeepSeek-V3 and ChatGPT-4. Meanwhile, the Advisors can be equipped with self-evolution capabilities, leveraging distillation techniques to develop a specialized model based on DeepSeek-R1, enabling more precise assessments of the communication environment.
\begin{figure*}
    \centering
    \includegraphics[width=0.95\linewidth]{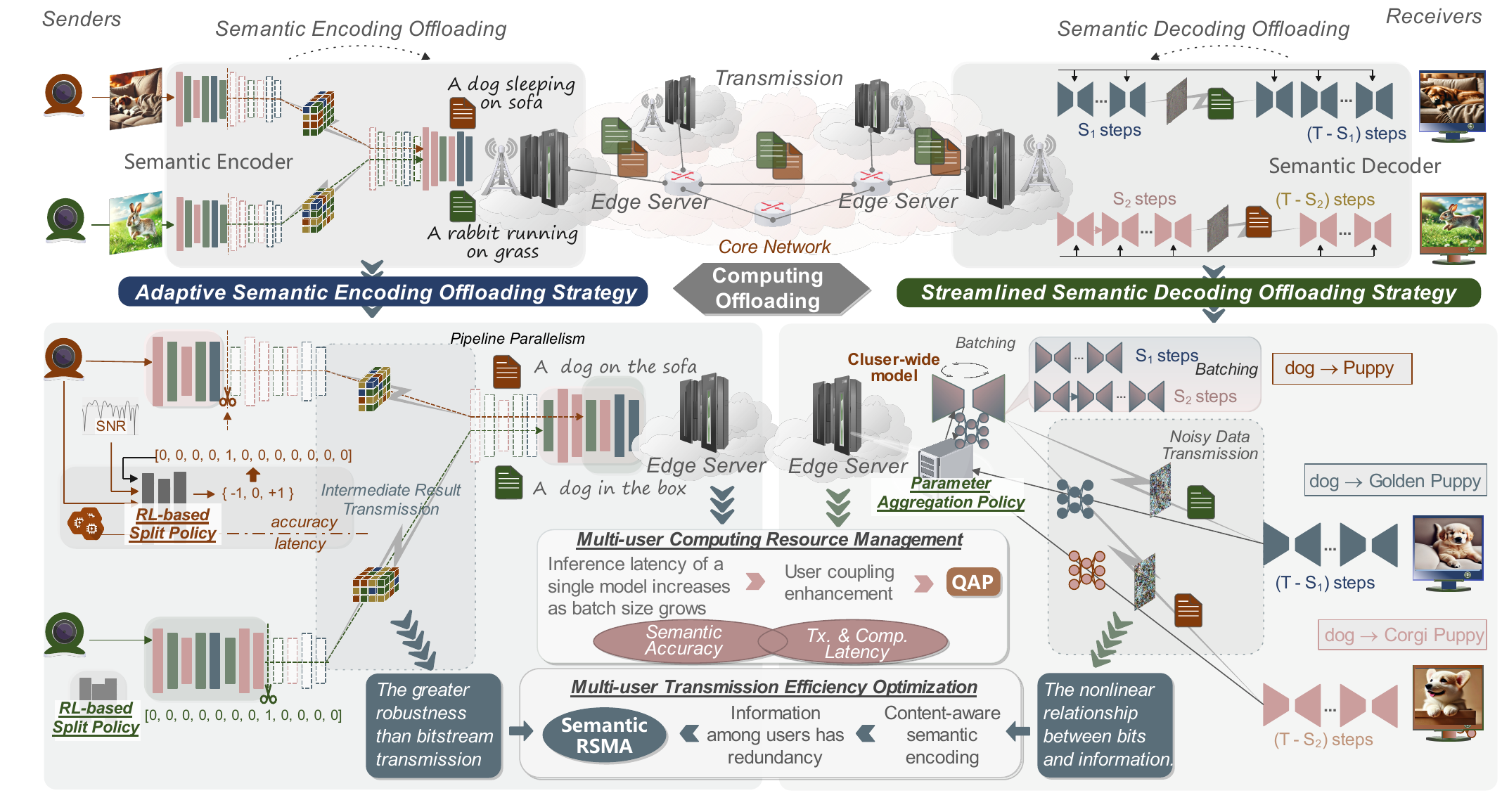}
    \caption{Optimization on offloading strategies for semantic encoding and decoding and issues in multi-user resource management.}
    \label{fig:offloading}
\end{figure*}
\subsection{Edge-Assisted Semantic Offloading Optimization}
The decoupled nature of generative SemCom results in the heterogeneity in semantic encoders and decoders. Thus, we propose two customized offloading paradigms for semantic encoding and decoding,  respectively.
\subsubsection{\textbf{Adaptive Offloading for Semantic Encoding}}
\label{sec: edge-encoder}
Although standardized semantic extraction models are increasingly lightweight, some heavy-weight models like SAM remain unsuitable for many devices.
A common solution is to partition the model between the end device and the edge, connected via a wireless link.
However, most offloading algorithms focus primarily on latency and energy consumption, overlooking the impact of wireless channel variability on inference accuracy. Fortunately, techniques of DL-based SemCom, such as channel encoder-decoder design or trainable quantization~\cite{guo2023digital}, can enhance semantic encoding offloading reliability. 
Moreover, a trainable feature distortion evaluation module can be designed  based on the Kullback-Leibler divergence~\cite{hu2024semharq} between the transmitted and received signals. It can assess the distortion levels of offloaded features and  determine whether the current offloading round is feasible. 
Nonetheless, due to the diverse computational capabilities of end devices and fluctuating edge resources, fixed offloading strategies often prove ineffective, underscoring the need for adaptive encoding offloading.

Considering these factors, reinforcement learning (RL) offers an effective approach to adaptive offloading. It enables the  integration of both wireless environment dynamics and computational resource variations into the RL paradigm as part of the environment.
However, existing RL-based adaptive offloading studies ~\cite{chen2024adaptive} overlook the limitation of available computational resources.  This is crucial in multi-user offloading scenarios,  at both the end and at the edge. Moreover, experimental results in~\cite{guo2023digital} indicate that even when computational resources remain constant, the optimal offloading layer varies depending on the raw data input.  To address these challenges, we propose an RL-based encoding offloading strategy.
We assume that both the sender's local devices and the edge server deploy a pre-trained semantic encoding model. 
As illustrated in  Fig.~\ref{fig:offloading}, during the Markov decision process (MDP) modeling, we introduce a new concept: the split point, which partitions  the entire network into two parts. The input-side layers are processed locally on the user device, while the output-side layers are executed on the edge server.
Specifically, the observable states in this MDP model include the source data, channel conditions, available computational resources at end and edge, and the current split point. The overall encoding latency and final communication performance serve as reward signals, guiding the agent to dynamically adjust the split point based on environmental changes. To facilitate the agent to capture sequential correlation between performance and different split points, we define the action space  as a relative adjustment from the current split point~\cite{chen2024adaptive}. Furthermore, during the RL training, the semantic encoder model can undergo alternating fine-tuning,  further enhancing the overall performance.

%

\label{sec: edge-decoder}

\subsubsection{\textbf{Efficient Offloading for Semantic Decoding}}
\label{sec: edge-decoder}
The \textit{recurrent} nature of diffusion models renders the offloading strategy described in Section~\ref{sec: edge-encoder} impractical. Splitting the neural network between  edge and end device introduces significant communication overhead during both training and inference. Therefore, to mitigate decoding latency,  it is essential to adopt an offloading strategy that splits the iterative inference process in temporal domain. 
Recent simulations indicate that the initial denoising steps primarily capture low-level features, such as contours and blobs, while personalized features emerge only in the later stages of the denoising process. This characteristic allows the initial denoising steps to generalize more effectively across similar tasks or personalized users. Additionally, the capability of diffusion models to capture multimodal distributions enables them to handle heterogeneous data more efficiently. Leveraging this property, we propose deploying a \textit{cluster-wide} model at the edge  to facilitate decoding offloading for a user cluster with similar tasks, as illustrated in Fig.~\ref{fig:offloading}. 

Specifically, in the proposed strategy, the complete denoising process is split into two phases, called general denoising process and personalized denoising process, respectively.  In M-GSC framework, the semantic information from multiple users is first routed to the cluster-wide model at the edge server, near the receivers, where the initial general denoising process is performed.  Subsequently, the edge server forwards the intermediate noisy data, along with the semantic information, to the respective receivers. 
At last, each receiver  utilizes its customized local decoder to execute the latter personalized denoising process.
Given that receivers may be unwilling to share private data for training the cluster-wide model, a federated learning (FL)-based collaborative mechanism can be employed. In this approach, each user trains a personalized decoder on their local dataset and periodically uploads the results to the edge server for parameter aggregation and cluster-wide model updates.
In this framework, the parameter aggregation method plays a crucial role. Techniques such as adaptive weighting, gradient clipping, and normalization ensure refined integration of features from all users. This helps the cluster-wide model capture common semantic features from the data generated by personalized semantic decoders while maintaining personalization of the local models.
In scenarios where users have insufficient local data, training local decoders may also rely on FL. In this context, a hierarchical clustering-based aggregation method becomes essential for efficiently coordinating the joint training of the shared cluster-wide model and personalized local models.


\subsection{Multi-user Resource Coordination Optimization  }
SemCom is designed to utilize end-device computing to reduce network transmission burden. This implies that computational latency is just as critical as communication latency. Therefore, we propose two tailored optimization strategies for  communication and computational resources, respectively, within the M-GSC framework.

 \begin{figure}
    \centering
    \includegraphics[width=1\linewidth]{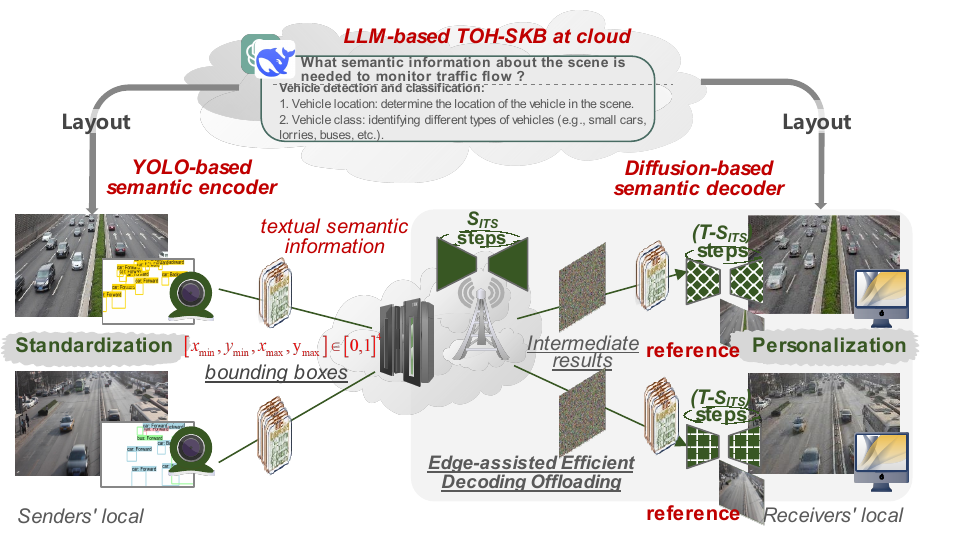}
    \caption{A simple M-GSC prototype for remote monitoring, where static scene backgrounds serve as personalized reference images.
 }
    \label{fig:prototype}
\end{figure}

\subsubsection{\textbf{Multi-user Transmission Efficiency Optimization}}
The M-GSC framework addresses  key challenges in the existing SemCom paradigm for multi-user communication tasks, such as large-scale end-to-end training and multi-relay forwarding in complex environments. It achieves this by decoupling multiple semantic encoders and decoders across users. However, this decoupling may introduce overlapping semantic information among the senders. For instance, in XR applications, the content captured by multiple cameras and sensors from different viewpoints inevitably contains redundancy data.
To mitigate this redundancy, integrating multi-user SemCom extraction design with rate-splitting multiple access (RSMA)   as a step toward semantic-aware semantic splitting multiple access (SSMA) presents a promising approach.

RSMA is originally a physical-layer technique that overlooks the content of individual user data streams. In RSMA, users share part of their data stream as a common message to improve transmission efficiency, but this comes at the cost of reduced privacy~\cite{jagatheesaperumal2024rate}. However, in content-aware SemCom systems, the determination of the common message can be redefined. Specifically, it can be derived from the redundant semantic information present in the data to be transmitted by multiple senders.
In scenarios requiring semantic encoding offloading, a distributed common feature extraction module can be deployed alongside distinctive feature extraction modules to enhance semantic encoder design. 
These modules can be jointly trained at the user-side encoders.
While this design partially reverts to the joint training paradigm of DL-SemCom, introducing higher training complexity and challenges in adaptive offloading, the use of analog SSMA-based transmission can significantly enhance semantic transmission capacity.
 Meanwhile, for scenarios with low computational requirements for semantic decoders, such as YOLO, a centralized control module can be added near the senders. This module processes explainable semantic information from individual sensors and extracts common information before transmission. Furthermore, the shared information extraction process can be jointly optimized with the MIMO beamforming matrix, improving overall system efficiency.

\subsubsection{\textbf{Multi-user Computing Resource Management} }
Given the dynamic variations in edge server resources and the heterogeneous computing power of terminal devices, dynamic optimization for offloading decision is crucial.
Uniquely, in M-GSC, offloading more denoising steps within a cluster-wide model can reduce decoding latency but also negatively impact the final personalized performance. Therefore, a trade-off metric balancing accuracy and latency must be carefully defined to guide the multi-user offloading optimization. Additionally, according to NVIDIA's report\footnote{https://www.nvidia.com/content/dam/en-zz/Solutions/Data-Center/tesla-product-literature/t4-inference-print-update-inference-tech-overview-final.pdf}, batching techniques cause inference latency to increase with batch size, typically following a linear function with a first-order term. Consequently, the response of one user's  offloading request can affect the overall latency reduction experienced by others. The increased interdependence among the receivers implies that the optimization objective becomes a quadratic function of multi-user offloading request response decisions.
Moreover, optimizing offloading steps for receivers granted offloading permissions transforms this multi-user offloading strategy into an extended generalized quadratic assignment problem.  Given the low complexity and high performance of  RL in solving combinatorial optimization problems, reformulating the original problem as an MDP decision sequence, where each user’s offloading decision is determined step by step, emerges as a promising approach. A prior study on this method can be found in~\cite{yang2024efficient}.

\begin{figure*}
    \centering
    \includegraphics[width=1\linewidth]{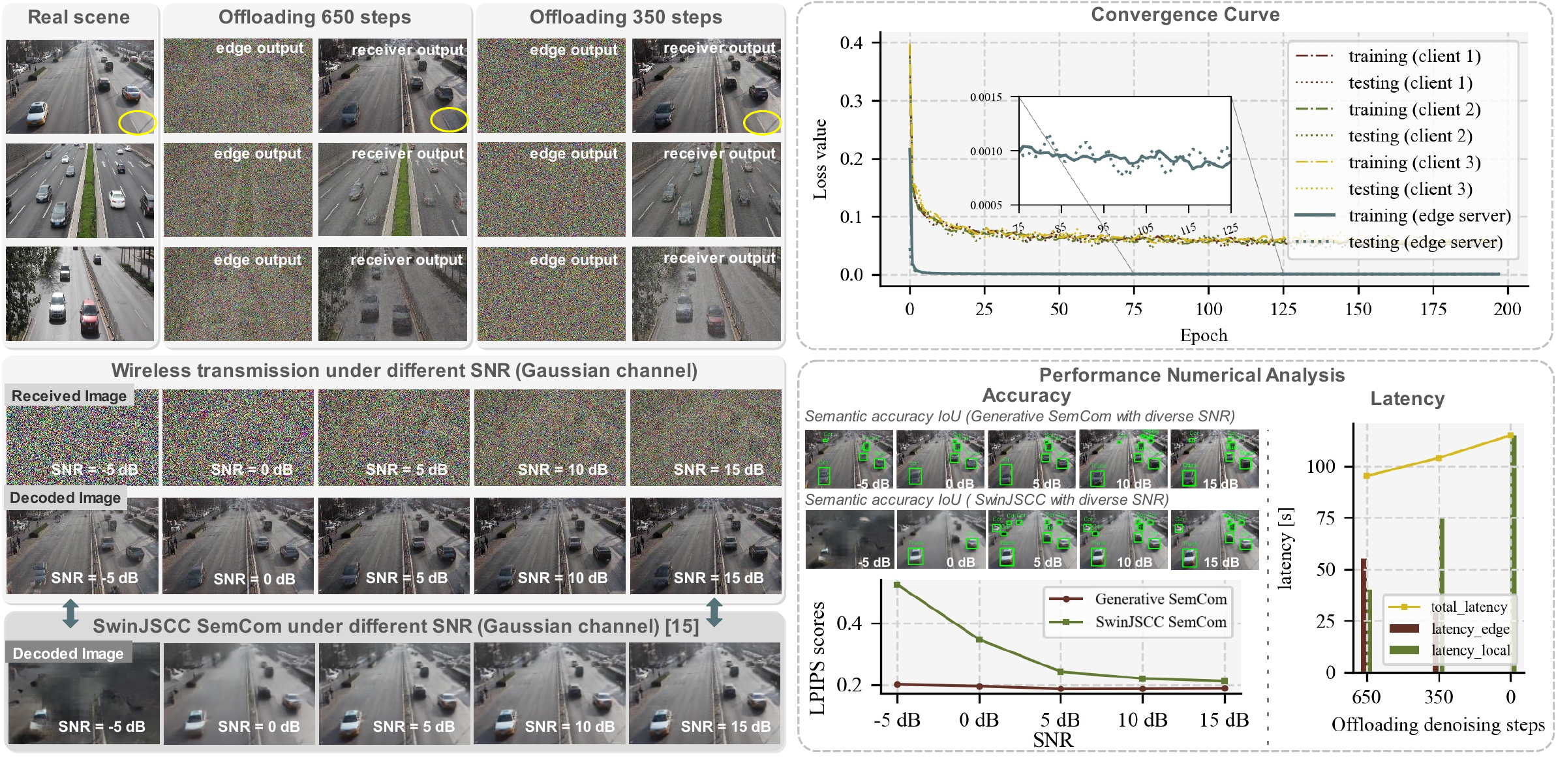}
    \caption{Simulation results for the streamlined decoding offloading. The simulations for edge server and receivers' local
are conducted on the computer equipped with NVIDIA A100-SXM4-80GB and  GeForce-RTX-4070-Ti-16G, respectively. The semantic evaluation results are visualized using YOLOv5.} 
    \label{fig:results}
\end{figure*}

In the above discussion, we only focus on single-pass communication. However, similar to traditional communication, error detection and retransmission mechanisms play a crucial role in enhancing reliability. The  difference lies in the increased inter-layer coupling in SemCom, which makes semantic transmission more robust than bitstream transmission. Therefore, a novel error detection and retransmission mechanism tailored for SemCom needs to be designed.

\section{Case Study}
\label{sec: casestudy}
In this section, we envision a simple multi-user SemCom prototype as shown in Fig.~\ref{fig:prototype}. 
We consider three bidirectional road traffic scenarios and select 1000 images from the UA-DETRAC dataset\footnote{https://detrac-db.rit.albany.edu/} for  three individual semantic decoder training. Additionally, 300 images per scenario is combined to create a dataset for the cluster-wide model training.
The maximum offloading steps are set to 650, with a total of 1,000 denoising steps\footnote{The classic DDPM is used in this case study. Advanced diffusion models, such as stable diffusion, which reduce denoising steps during inference, offer potential for broader applications.}. 
The semantic information of a monitored scene consists of multiple bounding boxes, each  representing an individual vehicle as a 4-tuple $\left(x_{\min }, y_{\min }, x_{\max }, y_{\max }\right)$.  Each element is normalized to the range $\left[0,1\right]$, and  encoded into five bits with a resolution of $0.03125$. Consequently, for a scene with $M$ vehicles, the total number of bits required is  $4 \times 5 \times M$.   We employ the digital transmission for semantic information due to its inherent digital existence and assume distortion-free transmission. Meanwhile, intermediate results are transmitted using highly efficient analog transmission, treating each pixel as a transmission symbol.

The convergence curves in Fig.~\ref{fig:results} demonstrate that, despite the high heterogeneity of the data used to train the shared model, the loss converges faster and to a lower value compared to the receivers' local loss. This phenomenon can be attributed to the fact that during the initial denoising phase, the model primarily captures  low-level features. This observation  supports the feasibility and effectiveness  of using a shared offloading model for similar tasks.
Additionally, Fig.~\ref{fig:results} contrasts the performance of offloading  350 steps and 650 steps to the edge server. The results indicate that both approaches yield satisfactory performance, though offloading 350 steps generally outperforms offloading 650 steps, albeit at the cost of longer latency, as shown in Fig.\ref{fig:results}. This suggests a trade-off between  computational latency and  the   quality of generated data.

Moreover, we simulate  the impact of wireless transmission between the edge server and receivers under different SNR levels with a Gaussian channel model. 
Meanwhile, we utilize the SwinJSCC~\cite{yang2024swinjscc}, a DL-based SemCom method,  as a benchmark algorithm to  evaluate its capability in generating \textit{photorealistic} representations. Moreover, with the analog transmission, the random noise in the wireless channel can be approximated as the forward noise addition process in the diffusion model. This similarity enhances the robustness of the proposed framework during the denoising process.
For this case study, we use Intersection over Union (IoU) to evaluate semantic accuracy and learned perceptual image patch similarity (LPIPS) to assess image quality, as it closely aligns with human perception. Before performance analysis, brightness equalization is applied to the decoded images. The results  indicate that generative SemCom produces higher-quality decoded images compared to SwinJSCC SemCom. Notably, under low SNR conditions, generative SemCom achieves significantly higher semantic accuracy than SwinJSCC.

\section{Conclusion}
\label{sec: con}
Based on the decoupling of semantic encoders and decoders in generative SemCom, we rethought its application in multi-user systems and proposed a novel M-GSC framework with an LLM-based SKB. Specifically,  we proposed a multi-LLM agent framework for enhanced SKB with greater robustness and adaptability. Moreover, we proposed two novel offloading paradigms designed specifically for adaptive semantic encoding offloading and efficient semantic decoding. Meanwhile, we highlighted tailored optimization strategies for the unique management of communication and computational resources in M-GSC. 
Finally, the effectiveness of the M-GSC framework was preliminarily validated through a case study focusing on efficient decoding offloading.

\section*{Acknowledgement}
The research is supported by the National Research Foundation, Singapore and Infocomm Media Development Authority under its Future Communications Research \& Development Programme. The research is also supported by the SUTD-ZJU IDEA Grant (SUTD-ZJU (VP) 202102), and the Ministry of Education, Singapore, under its Academic Research Fund Tier 2 (MOE-T2EP20221-0017), SMU-SUTD Joint Grant (22-SIS-SMU-048), and SUTD Kickstarter Initiative (SKI 20210204).

\bibliographystyle{IEEEtran}
\bibliography{ref}
\section*{Biographies}
\small
{WANTING YANG} is currently a research fellow and scientist with Singapore University of Technology and Design. She received the B.S. degree and the Ph.D. degree from the Department of Communications Engineering, Jilin University, Changchun, China, in 2018 and 2023, respectively. Her research interests include  semantic communication, deep reinforcement learning, martingale theory, edge computing, generative AI.

{ZEHUI XIONG} is currently with Singapore University of Technology and Design, Singapore. He received the PhD degree in Nanyang Technological University (NTU), Singapore. His research interests include wireless communications, Internet of Things, blockchain, edge intelligence, and Metaverse.  He has served as the Associate Director of Future Communications R\&D Programme, and Deputy Lead of AI Mega Centre.

{SHIWEN MAO} [IEEE Fellow] is a Professor and Earle C. Williams Eminent Scholar at Auburn University, Auburn, AL. He received his Ph.D. in electrical and computer engineering in 2004 from Polytechnic
University, Brooklyn, NY.   His research interest includes wireless networks and multimedia communications. He is the Editor-in-Chief of IEEE Transactions on Cognitive Communications and Networking.

{TONY Q. S. QUEK} [IEEE Fellow] is the Cheng Tsang Man Chair Professor, ST Engineering Distinguished Professor, and Head of ISTD Pillar with Singapore University of Technology and Design as well as the Director of Future Communications R\&D Programme. He was honored with the 2020 IEEE Communications Society Young Author Best Paper Award, the 2020 IEEE Stephen O. Rice Prize, the 2020 Nokia Visiting Professor, and the 2022 IEEE Signal Processing Society Best Paper Award. He is a Fellow of IEEE and a Fellow of the Academy of Engineering Singapore.

{PING ZHANG} [IEEE Fellow] is currently a Professor with Beijing University of Posts and Telecommunications, the Director of the
State Key Laboratory of Networking and Switching Technology and the Department of Broadband Communication of Peng Cheng Laboratory, and a member of IMT-2020 (5G) Experts Panel and Experts Panel for China’s 6G Development. He was the Chief Scientist of the National Basic Research Program (973
Program). His research interests include semantic communication.
He is an Academician of Chinese Academy of Engineering.

\MakeUppercase{Mérouane Debbah} is a Professor at Khalifa University of Science and Technology in Abu Dhabi and founding Director of the KU 6G Research Center. He is a frequent keynote speaker at international events in the field of telecommunication and AI. He is known for his work on Large Language Models, distributed AI systems for networks and semantic communications. He received multiple prestigious distinctions, prizes and best paper awards for his contributions to both fields. He is a WWRF Fellow, a Eurasip Fellow, an AAIA Fellow, an Institut Louis Bachelier Fellow and a Membre émérite SEE.

\MakeUppercase{Rahim Tafazolli},  CBE, FIEEE is a Regius Professor of Electronic Engineering, Professor of  mobile and satellite communications and the Director of the Institute for Communication Systems (ICS), University of Surrey, Guildford, U.K. He is also the Founder of the 5G and 6G Innovation Centre (5GIC and 6GIC), University of Surrey. He was elected as a fellow of the U.K. Royal Academy of Engineering (FREng)  in 2020 and FIET, FCIC and a Fellow of Wireless World Research Forum (WWRF).

\end{document}